\documentstyle[12pt]{article}
\def\thebibliography#1{\bigskip\section*{\centering
References\\}\bigskip\list
{\arabic{enumi}.}{\settowidth\labelwidth{#1}\leftmargin\labelwidth
\advance\leftmargin\labelsep
\usecounter{enumi}}
\def\newblock{\hskip .11em plus .33em minus .07em}
\sloppy\clubpenalty4000\widowpenalty4000
\sfcode`\.=1000\relax}

\def\op#1{\mathop{\fam0 #1}\limits}

\newcommand{\ben}{\begin{eqnarray}}
\newcommand{\een}{\end{eqnarray}}

\newcommand{\be}{\begin{eqnarray*}}
\newcommand{\ee}{\end{eqnarray*}}
\newcommand{\bea}{\begin{eqalph}}
\newcommand{\eea}{\end{eqalph}}
\newcommand{\cL}{{\cal L}}
\newcommand{\bL}{{\bf L}}
\newcommand{\R}{{\bf R}}
\newcommand{\C}{{\bf C}}

\newcommand{\cE}{{\cal E}}

\newcommand{\cF}{{\cal F}}

\newcommand{\cD}{{\cal D}}

\newcommand{\al}{\alpha}
\newcommand{\bt}{\beta}
\newcommand{\kp}{\kappa}
\newcommand{\dl}{\delta}
\newcommand{\la}{\lambda}

\newcommand{\ap}{\approx}

\newcommand{\f}{\phi}
\newcommand{\om}{\omega}
\newcommand{\ot}{\otimes}
\newcommand{\Om}{\Omega}
\newcommand{\m}{\mu}
\newcommand{\n}{\nu}
\newcommand{\g}{\gamma}
\newcommand{\G}{\Gamma}
\newcommand{\e}{\epsilon}
\newcommand{\ve}{\varepsilon}

\newcommand{\Si}{\Sigma}
\newcommand{\si}{\sigma}

\newcommand{\w}{\wedge}

\newcommand{\wt}{\widetilde}
\newcommand{\wh}{\widehat}
\newcommand{\ol}{\overline}
\newcommand{\dr}{\partial}

\newcounter{eqalph}
\newcounter{equationa}

\newenvironment{eqalph}{\stepcounter{equation}
\setcounter{equationa}{\value{equation}}
\setcounter{equation}{0}

\begin{eqnarray}}{\end{eqnarray}
\setcounter{equation}{\value{equationa}}}

\begin{document}
\hbox{}

\centerline{\large\bf Stress-Energy-Momentum Tensors}
\medskip

\centerline{\large\bf in Lagrangian Field Theory.}
\bigskip

\centerline{\large\bf Part 2. Gravitational Superpotential.}
\bigskip

\centerline{\sc Giovanni Giachetta}

\medskip

\centerline{Department of Mathematics and Physics}

\centerline{University of Camerino, 62032 Camerino, Italy}

\centerline{E-mail: mangiarotti@camvax.unicam.it}
\medskip

\centerline{\sc Gennadi Sardanashvily}
\medskip

\centerline{Department of Theoretical Physics}

\centerline{Moscow State University, 117234 Moscow, Russia}

\centerline{E-mail: sard@grav.phys.msu.su}
\bigskip

\begin{abstract}
Our investigation of differential conservation laws in Lagrangian field theory
is based on the first variational formula which provides the canonical
decomposition of the Lie  derivative of a Lagrangian density by a projectable
vector field on a bundle (Part 1: gr-qc/9510061). If a Lagrangian density is
invariant under a certain class of bundle isomorphisms, its Lie derivative by
the associated vector fields vanishes and the corresponding differential
conservation laws take place. If these vector fields depend on derivatives of
parameters of bundle transformations, the conserved current reduces to a
superpotential. This Part of the work is devoted to gravitational
superpotentials. The invariance of a gravitational Lagrangian density under
general covariant transformations leads to the stress-energy-momentum
conservation law where the energy-momentum flow of gravity reduces to the
corresponding generalized Komar superpotential. The associated energy-momentum
(pseudo) tensor can be defined and calculated on solutions of metric and
affine-metric gravitational models.
\end{abstract}
 \newpage

{\bf CONTENTS}
\bigskip

\noindent
1. Discussion \newline
2. Geometric Preliminary
\medskip

{\sc Part 1.}
\newline
3. Lagrangian Formalism of Field Theory\newline
4. Conservation Laws\newline
5. Stress-Energy-Momentum Conservation Laws\newline
6. Stress-Energy Conservation Laws in Mechanics \newline
7. Noether Conservation Laws \newline
8. General Covariance Condition\newline
9. Stress-Energy-Momentum Tensor of Matter Fields \newline
10. Stress-Energy-Momentum Tensors of Gauge Potentials \newline
11. Stress-Energy-Momentum Tensors of Proca Fields \newline
12. Topological Gauge Theories
\medskip

{\sc Part 2.}
\newline
13. Reduced Second Order Lagrangian Formalism\newline
14. Conservation Laws in Einstein's Gravitation Theory\newline
15. First Order Palatini Formalism\newline
16. Energy-Momentum Superpotential of Affine-Metric Gravity\newline
17. Lagrangian Systems on Composite Bundles\newline
18. Composite Spinor Bundles in Gravitation Theory\newline
19. Conservation Laws in the Gauge Gravitation Theory
\newpage

\centerline{\large \bf PART 2}
\bigskip \bigskip

The present Part of the work is devoted to SEM conservation laws and
energy-momentum superpotentials in the gravitation theory.

There are different approaches to the description of conservation laws in the
gravitation theory; different
 energy-momentum (pseudo) tensors of gravity have been suggested
\cite{vir,bak,chi,heh,kom,reg,sza}. Our analysis of the energy-momentum
conservation law in the gravitation theory is based on the first variational
formula in Lagrangian field theory (see Sections 3, 4 of the Part 1
\cite{cam1}). In accordance with this formula, the invariance of a
gravitational Lagrangian density under general covariant transformations
implies the differential SEM conservation law where the corresponding
energy-momentum flow reduces to the superpotential \cite{bor,fer10,cam2,nov}.

Let us briefly remind the basic features of the geometric approach to field
theory when classical fields are described by global sections of a bundle
$Y\to X$ over a world manifold $X$ and their dynamics is phrased in terms of
jet manifolds (see Section 3).

We restrict ourselves
to the first order  Lagrangian formalism, for
 most of contemporary field models are described by
first order Lagrangian densities. This is not the case for
 the Einstein-Hilbert Lagrangian density of the Einstein's
gravitation theory which belongs to the special class of second order
Lagrangian densities. Its Euler-Lagrange equations are however of the order
two as like as in the first order theory (see Sections 13, 14).

In the first order Lagrangian formalism, the finite-dimensional
configuration space
of fields represented by sections $s$ of a bundle $Y\to X$ is the first order
jet manifold $J^1Y$ of $Y$.  Given fibered coordinates $(x^\m,y^i)$
of $Y$, the jet manifold $J^1Y$ is endowed with the adapted  coordinates
$ (x^\m,y^i,y^i_\m)$.
A first order
Lagrangian density on $J^1Y$ is defined to be an exterior horizontal density
$$
L=\cL(x^\m,y^i,y^i_\m)\om, \qquad \om=dx^1\w ...\w dx^n.
$$

Let $G_t$ be a 1-parameter group of bundle isomorphisms of a bundle $Y\to X$,
and let
$$
u= u^\la (x)\dr_\la + u^i(y)\dr_i
$$
be the corresponding projectable vector field on $Y$. A
Lagrangian density  $L$ on the configuration space $J^1Y$ is invariant under
these transformations iff its Lie derivative by the jet lift
$$
j^1_0u=u^\la\dr_\la + u^i\dr_i + (\wh\dr_\la u^i
- y_\m^i\dr_\la u^\m)\dr_i^\la,
$$
$$
\wh\dr_\la = \dr_\la +y^j_\la \dr_j+y^j_{\m\la}\dr_j^\m +\cdots
$$
of $u$ onto $J^1Y$ is equal to zero, i.e.
\begin{equation}
\bL_{j^1_0u}L=0. \label{2C2}
\end{equation}

In accordance with the first variational formula (35), there is the canonical
decomposition
\begin{equation}
\bL_{j^1_0u}L=  u_V\rfloor\cE_L+ dT(u) \label{2C3}
\end{equation}
where
\begin{equation}
\cE_L=
 [\dr_i-(\dr_\la +y^j_\la\dr_j+y^j_{\m\la}\dr^\m_j)
\dr^\la_i]\cL dy^i\w\om=\dl_i\cL dy^i\w\om \label{2C10}
\end{equation}
is the Euler-Lagrange operator,
$$
T(u)=T^\la(u)\om_\la =[\pi^\la_i(u^i-u^\m y^i_\m )+u^\la\cL]\om_\la,
$$
$$
\pi^\m_i=\dr^\m_i\cL,
\qquad \om_\la =\dr_\la\rfloor\om,
$$
is the corresponding current,
and
$$
u_V=(u^i -y^i_\m u^\m)\dr_i
$$
is the vertical part of the vector field $u$.

The
Euler-Lagrange operator  $\cE_L$, by definition, vanishes on the critical
sections of the bundle $Y\to X$, and the equality (\ref{2C3}) comes to the weak
identity
$$
 {\bf L}_{j^1_0u}L\ap \wh\dr_\la[\pi^\la_i(u^i- u^\m
y^i_\m) +u^\la\cL ]\om
$$
(the symbol "$\ap $" means equality modulo the kernel of the Euler-Lagrange
operator (\ref{2C10})).

If the Lie derivative
$$
\bL_{j^1_0u}L=[\dr_\la u^\la\cL +(u^\la\dr_\la+
u^i\dr_i +(\dr_\la u^i +y^j_\la\dr_ju^i -y^i_\m\dr_\la u^\m)\dr^\la_i)\cL]\om
$$
of a Lagrangian density $L$ by a projectable vector field $u$ satisfies the
condition (\ref{2C2}), then
we get the weak conservation law
\begin{equation}
0\ap \wh \dr_\la[\pi^\la_i(u^i-u^\m y^i_\m )+u^\la\cL]. \label{K4}
\end{equation}
Note that, if some background fields $y^A$ are present, the corresponding
variational derivatives $\dl_A\cL$ in the Euler-Lagrange operator (\ref{2C10})
do not vanish, and we have the differential transformation law
\begin{equation}
0\ap (u^A- u^\m y^A_\m)\dl_A\cL +
\wh \dr_\la[\pi^\la_A(u^A-u^\m y^A_\m ) +
\pi^\la_i(u^i-u^\m y^i_\m )+u^\la\cL] \label{2C11}
\end{equation}
(see Section 4).

The examples of gauge fields (see Section 7) and tensor fields (see Section
8) show that, in case of gauge-type transformations when the corresponding
vector fields $u$ depend on derivatives of the parameters $\al(x)$ of these
transformations, the conservation law (\ref{K4}) takes the form
$$
T=W+dU
$$
where $W\ap 0$ and $U$ is a superpotential which depends on parameters $\al(x)$
of gauge transformations that provide the gauge invariance of the
conservation law (\ref{K4}).

The Einstein's gravitation
theory and the affine-metric gravitation theory are field models on
the bundles of geometric objects (see Section 8). Gravitational Lagrangian
densities are invariant under general covariant transformations. As a
consequence, we get the SEM conservation law with respect to the canonical
lift of a vector field $\tau$ on a world manifold $X$ onto the
corresponding bundle of gravitational fields.

In the purely metric Einstein's gravitation theory
\cite{nov}, the SEM conservation law
 corresponding to the invariance of the Hilbert-Einstein
Lagrangian density under general covariant transformations takes the form
\begin{equation}
\frac{d}{dx^\la}T^\la (\tau)\ap  0, \qquad  T^\la \ap\frac{d}{dx^\la}U^{\m\la}
(\tau),\label{K1}
\end{equation}
where
\begin{equation}
U^{\m\la}(\tau) =\frac{\sqrt{-g}}{2\kp}
( g^{\la\nu}\tau^\m_{;\nu} -g^{\m\nu}\tau^\la_{;\nu})  \label{K2}
\end{equation}
is the well-known Komar superpotential \cite{kom} associated with a vector
field $\tau$ on a world manifold $X$. Here the symbol "$_{;\m}$" denotes the
covariant derivative with respect to the Levi-Civita connection.

In the recent paper \cite{bor}, it was shown that (\ref{K3}) has a kind of
universal property, in the sense that the SEM flow of any Lagrangian density
depending nonlinearly on the scalar curvature, constructed from a metric and
a torsionless connection, reduces always to the Komar superpotential.

This result has been extended to the affine-metric gravity in case of a general
linear connection $K^\al{}_{\g\m}$ and arbitrary Lagrangian density $L$
invariant under general covariant transformations  \cite{cam2}. The
corresponding SEM conservation law is brought into the form (\ref{K1}) where
\begin{equation}
U^{\m\la}(\tau) =\frac{\dr\cL}{\dr K^\al{}_{\nu\m,\la}} (D_\nu
\tau^\al + \Om^\al{}_{\nu\si}\tau^\si)  \label{K3}
\end{equation}
is the generalized Komar superpotential. Here $D_\g$ is
the covariant derivative with respect to the general linear connection
$K$ and $\Om$ is the torsion of this connection. In the particular case of
the Hilbert-Einstein Lagrangian density and symmetric connections, we have
$$
\frac{\dr\cL_{\rm HE}}{\dr K^\al{}_{\nu\m,\la}} =
\frac{\sqrt{-g}}{2\kp}
(\dl^\m_\al g^{\nu\la} - \dl^\la_\al g^{\nu\m}),
$$
so that the superpotential (\ref{K3}) comes to the Komar
superpotential (\ref{K2}).
Also, if the Lagrangian density is of the
kind considered in Ref.\cite{bor},
the superpotential (\ref{K3}) recovers the superpotential found in that paper.

In case of also gauge gravitation theory, we show that the covariant
derivative of Dirac fermion fields takes the form
\begin{equation}
\wt D_\la  =\dr_\la -\frac12A^{ab}{}_\m^c (\dr_\la h^\m_c +
K^\m{}_{\nu\la}  h^\nu_c)I_{ab}, \label{K101}
\end{equation}
$$
A^{ab}{}_\m^c =\frac12(\eta^{ca}h^b_\m -\eta^{cb}h^a_\m),
$$
where $h(x)$ is a tetrad gravitational field, $K$ is a general linear
connection, $\eta$ is the Minkowski metric, and
 $$
I_{ab}=\frac14[\g_a,\g_b]
$$
are the generators of the spinor group $L_s= SL(2,\C)$ \cite{cam3}.

The covariant derivative (\ref{K101}) has been considered by several authors
\cite{ar,pon,tuc}. In accordance with the
well-known theorem \cite{kob}, every general linear connection being
projected onto the Lie algebra of the Lorentz group yields a Lorentz
connection.

It follows that the configuration
space of metric (or tetrad) gravitational fields and general linear
connections  may play the role of the universal configuration space of
realistic gravitational models. In particular, one can think of the
generalized Komar superpotential as being the universal  energy-momentum
superpotential of gravity. The corresponding energy-momentum (pseudo) tensor
reads
$$
T^\la_\al=\frac{d}{dx^\m}(\frac{\dr\cL}{\dr K^\si{}_{\nu\m,\la}}
K^\si{}_{\nu\al}).
$$
One can calculate it on solutions of metric and affine-metric gravitational
models. In particular, the torsion contributes into this energy-momentum
(pseudo) tensor.

Note that the dependence of the energy-momentum superpotentials
of gravity on the vector field $\tau$ reflects the fact that the SEM
conservation law  (\ref{K1}) is preserved under general covariant
transformations.

Hereafter, the 4-dimensional base manifold $X$ is required to satisfy the
well-known topological condition in order that a pseudo-Riemannian  metric can
exist. To
summarize these conditions, we assume that the manifold $X$ is not compact and
that the tangent bundle of $X$ is trivial.
 We call $X$ the world manifold. Pseudo-Riemannian metrics and
general linear connections in tangent and cotangent bundles of $X$ are called
the world metrics and the world connections respectively.
\bigskip\bigskip

\noindent
{\Large \bf 13\,\,\, Reduced second order Lagrangian formalism}
\bigskip\bigskip

Given a bundle $Y\to X$ coordinatized by
$(x^\la, y^i)$, let $L$ be a second order Lagrangian density
on the second order jet manifold $J^2Y$ of $Y$. Its different Lepagian
equivalents exist on $J^3Y$, but the associated
Poincar\'e-Cartan form $\Xi_L$ is uniquely defined and given by the
coordinate expression
\begin{equation}
\Xi_L=\cL\om +[(\dr^\la_i\cL -\wh\dr_\m\dr_i^{\la\m}\cL)\wh dy^i +
\dr_i^{\m\la}\cL\wh dy^i_\m]\w\om_\la. \label{C117}
\end{equation}

In the second order case, the first variational formula (23) is
written
$$
\pi^{4*}_2\bL_{j^2_0u}L=
 h_0(j^3_0u\rfloor d\rho_L) + h_0d(j^3_0u\rfloor\rho_L)
$$
for any projectable vector field  $u$ on the bundle $Y\to X$.
When $\rho_L=\Xi_L$, it takes the form
\begin{equation}
\pi^{4*}_2\bL_{j^2_0u}L
= u_V\rfloor \cE_L + d_H h_0(\ol u\rfloor\Xi_L) \label{C118}
\end{equation}
where
 \begin{equation}
\cE_L= (\dr_i-\wh\dr_\la\dr^\la_i +\wh\dr_\m\wh\dr_\la\dr^{\la\m}_i)\cL
dy^i\w\om \label{C119}
\end{equation}
is the 4-order Euler-Lagrange operator and the second order jet lift
$j^2_0u$ of the vector field $u$ reads
$$
j^2_0u=u^\la\dr_\la +u^i\dr_i +(\wh\dr_\la u^i -y^i_\nu\dr_\la u^\nu)
\dr^\la_i +[\wh\dr_\m(\wh\dr_\la u^i -y^i_\nu\dr_\la u^\nu) -y^i_{\la\nu}
\dr_\m u^\nu ]\dr_i^{\la\m}.
$$

Being restricted to the kernel of the Euler-Lagrange operator (\ref{C119}),
the equality (\ref{C118}) comes to the weak identity
\begin{equation}
\pi^{4*}_2\bL_{j^2_0u}L \ap \wh\dr_\la[u^\la\cL +
u_V{}^i(\dr^\la_i\cL -\wh\dr_\m\dr_i^{\la\m}\cL) +
\wh \dr_\m u_V{}^i\dr_i^{\m\la}\cL]\om \label{C122}
\end{equation}
and to the corresponding differential transformation law on critical sections
of the bundle $Y\to X$.

Let us consider a second order Lagrangian density $L$
whose Euler-Lagrange operator $\cE$ (\ref{C119}) reduces to the second order
differential operator \cite{kru82}. It takes place, if the
associated Poincar\'e-Cartan form  $\Xi_L$ (\ref{C117}) is defined on the first
order jet manifold $J^1Y$ of $Y$. This is the case iff the Lagrangian density
$L$ obeys the conditions
\ben
&& \dr_j^{\al\bt}\dr_i^{\m\nu}\cL =0, \nonumber\\
&& (\dr^\nu_j\dr_i^{\m\la} -\dr_i^\m\dr_j^{\nu\la})\cL=0. \label{C120}
\een
This Lagrangian density is linear in the coordinates $y^i_{\la\m}$ and,
in each coordinate
chart, it is given by the expression
\begin{equation}
L=(\cL' +\pi_i^{\m\la} y^i_{\m\la})\om \label{C121}
\end{equation}
where $\cL'$  is a local function on $J^1Y$ and the Lagrangian momentum
$\pi$ is a section of the Legendre bundle
$$
\op\w^nT^*X\op\otimes_{J^1Y}(\op\vee^2TX)\op\otimes_{J^1Y}V^*Y.
$$
In virtue of the relation (\ref{C119}), there exists a local horizontal
form $\f=\f^\la\om_\la$ on $J^1Y\to X$ such that
$$
\pi^{\m\la}_i =\dr^\m_i\f^\la.
$$
Let us consider the local form
$$
\e=\Xi_L -d\f.
$$
It is the Lepagian equivalent of the  local first order Lagrangian density
\begin{equation}
L_1=h_0(\e)=L-d_H\f\label{C123}
\end{equation}
which leads to the same second order Euler-Lagrange operator in a given
coordinate chart as the Lagrangian density (\ref{C121}) does.

In particular, if the functions $\pi_i^{\m\la}$ are independent of the
coordinates $y^i_\m$, we can take
\begin{equation}
\f =\pi_i^{\m\la}(y^i_\m -\G^i_\m)\om_\la \label{C124}
\end{equation}
where $\G$ is a connection on $Y\to X$. The form (\ref{C124}) is globally
defined and we get the first order Lagrangian density
\begin{equation}
L_1 =L- \wh\dr_\la[\pi_i^{\m\la}(y^i_\m -\G^i_\m)]\om \label{C125}
\end{equation}
which leads to the same second order Euler-Lagrange operator as the
Lagrangian density (\ref{C121}) does.

 However, the
first order Lagrangian densities $L_1$ (\ref{C123}) and (\ref{C125}) fail to
possess the same symmetries of the second order Lagrangian density $L$
(\ref{C121}) in general. Therefore, we do not take advantage of its
application in conservation laws as a rule.
\bigskip\bigskip

\noindent
{\Large \bf 14\,\,\, Conservation laws in Einstein's gravitation theory
}
\bigskip\bigskip

In Einstein's gravitation theory, gravity is described by a pseudo-Riemannian
metric whose Lagrangian density is the Hilbert-Einstein Lagrangian density.

Let $\Si_g\to X$ be the bundle of pseudo-Riemannian world metrics. Its
2-fold covering is the bundle
\begin{equation}
\Si=LX/SO(3,1) \label{C126}
\end{equation}
where $LX$ is the principal bundle of oriented linear frames and $SO(3,1)$ is
the connected Lorentz group. Hereafter, we shall identify $\Si_g$ with the open
subbundle of the tensor bundle
$$
\op\vee^2T^*X\to X.
$$
In induced coordinates of $T^*X$, the bundle $J^2\Si_g$ is coordinatized by
$(x^\la, g_{\al\bt}, g_{\al\bt\la}, g_{\al\bt\la,\m})$.

The second order Hilbert-Einstein Lagrangian density on the configuration space
$J^2\Si_g$ reads
\begin{equation}
L_{\rm HE}=-\frac{1}{2\kp}g^{\al\mu}g^{\bt\nu}R_{\al\bt\m\nu}
\sqrt{-g}\omega\label{C127}
\end{equation}
where
$$
R_{\al\bt\m\nu} =\frac12 (g_{\al\nu\bt\mu}+ g_{\bt\m\al\nu} -g_{\al\m\bt\nu}
-g_{\bt\nu\al\m}) + g^{\ve\si}(\g_{\ve\bt\m}\g_{\si\al\nu} -
\g_{\ve\bt\nu}\g_{\si\al\m}),
$$
$$
\g_{\al\m\nu} =\frac12( g_{\al\m\nu} +g_{\al\nu\m} -g_{\m\nu\al}),
$$
$$
g^{\al\bt} =\frac1g\frac{\dr g}{\dr g_{\al\bt}}, \qquad
\frac{\dr}{g_{\al\bt}} = -g^{\al\m}g^{\bt\nu}\frac{\dr}{g^{\m\nu}}.
$$
This is the reduced second order Lagrangian density like that
considered in the previous Section. It leads to the second order Euler-Lagrange
operator.

To remain within the framework of bundles of geometric
objects, we utilize the Proca fields as the matter source of gravitational
fields. Their Lagrangian density (90) depends on a world metric, but not on
the symmetric part of the world connection.

The Proca fields are described by sections of the cotangent bundle $T^*X$
(see Section 11). The total configuration space of metric gravitational fields
and Proca fields is $J^2T$ where
$$
T=\op\vee^2 T^*X\op\times_X T^*X.
$$
On this configuration space, the Lagrangian density of Proca fields is given
by the expression
\begin{equation}
L_{\rm P}=[-\frac{1}{4\g}g^{\mu\al}g^{\nu\beta}\cF_{\al
\beta}\cF_{\mu\nu}
 -\frac12 m^2g^{\mu\la}k_\mu k_\la]\sqrt{\mid g\mid}\omega, \label{C128}
\end{equation}
where
$$
\cF_{\mu\nu} = k_{\nu\mu} -k_{\mu\nu}
$$
and $g^{\nu\bt}$ is the inverse matrix of  $g_{\nu\bt}$.

The total Lagrangian density $L$ is the sum of the Hilbert-Einstein Lagrangian
density  (\ref{C127}) and the Lagrangian density of Proca fields  (\ref{C128}).

The associated Poincar\'e-Cartan form on the jet manifold
$J^1T$ reads
$$
\Xi_L=\Xi_{\rm HE} +\Xi_{\rm P}
$$
where
\be
&&\Xi_{\rm HE} =-\frac1{2\kp}\sqrt{-g}[g^{\al\m} g^{\bt\nu}g^{\ve\si}
(\g_{\ve\m\nu}\g_{\al\bt\si} -
\g_{\ve\bt\al}\g_{\m\si\nu})\om +\\
&&\qquad (g^{\m\al}g^{\la\bt}
-g^{\al\bt}g^{\m\la}) (dg_{\al\bt\m} +g^{\nu\si}\g_{\si\al\bt}
dg_{\m\nu})\w\om_\la]
\ee
\cite{kru82} and $\Xi_{\rm P}$ is given by the expression
\begin{equation}
\Xi_{\rm P}= (\cL_{\rm P}
 -\pi^{\m\la}k_{\mu\la})\om + \pi^{\m\la}dk_\mu\w\om_\la, \label{C137}
\end{equation}
$$
\pi^{\m\la}= -\frac1{\g}g^{\mu\al}g^{\la\beta}\cF_{\beta\al}\sqrt{\mid g\mid}.
$$
The total Euler-Lagrange operator is
$$
\cE_L =\cE_{\rm HE} +\cE_{\rm P},
$$
\ben
&&\cE_{\rm HE}= -\frac12g^{\al\m}g^{\bt\nu}\sqrt{-g}[-\frac1\kp(R_{\m\nu}
-\frac12g_{\m\nu}R) + t_{\m\nu}]dg_{\al\bt}\w\om
=\dl^{\al\bt}\cL dg_{\al\bt}\w\om \nonumber\\
&&
\cE_{\rm P}= [-\sqrt{-g} m^2g^{\al\bt}k_\al +\frac1\g\wh\dr_\al
(g^{\mu\al}g^{\nu\beta}\cF_{\mu\nu}\sqrt{-g})]dk_\bt\w\om=
 \dl^\bt\cL  dk_\bt\w\om,  \label{C129}
\een
where
$t$ is the metric energy-momentum tensor (94) of the Proca
fields.

The Lagrangian
densities $L_{\rm HE}$ (\ref{C127}) and $L_{\rm P}$
(\ref{C128}) are invariant separately under  general covariant
transformations of the bundle of geometric objects $T=\Si_g\op\times_X T^*X$.
Therefore, the Lie derivative of their sum $L$
by the jet lift
$j^2_0\wt\tau$ of
the vector field on $T$
$$
\wt\tau =\tau^\la\dr_\la - (g_{\nu\bt}\dr_\al\tau^\nu
+g_{\al\nu}\dr_\bt\tau^\nu)\frac{\dr}{\dr g_{\al\bt}} -\dr_\al\tau^\nu k_\nu
\frac{\dr}{k_\al}
$$
naturally induced by the vector field $\tau$ on $X$
vanishes. Hence, we get the week conservation law  which takes the
 form
\begin{equation}
0 \ap \wh\dr_\la[-2g_{\m\al}\tau^\m\dl^{\al\la}\cL
+ \tau^\nu k_\nu\dl^\la\cL
+\frac1{2\kp}\sqrt{-g}(g^{\m\nu}\tau^\la_{;\nu} -
g^{\la\nu}\tau^\m_{;\nu})_{;\m} -\wh\dr_\m(\pi^{\mu\la}\tau^\nu k_\nu)]
\label{C130}
\end{equation}
\cite{nov}.

A glance at the conservation law  (\ref{C130}) shows that the SEM flow of a
metric gravitational field with respect to the vector field $\wt\tau$
reduces to the  Komar superpotential (\ref{K2}). The total superpotential
contains also the superpotential
\begin{equation}
 Q^{\m\la}(\tau) =
\pi^{\mu\la}\tau^\nu k_\nu \label{2C13}
\end{equation}
of the Proca fields. We observe that, in case of exact general covariant
transformations, the energy-momentum flow (95) of the Proca fields comes to
the superpotential term (see Section 11).
\bigskip\bigskip

\noindent
{\Large \bf 15\,\,\,  First order Palatini formalism}
\bigskip\bigskip

This Section is devoted to the SEM conservation laws of
gravitational theory in the first order  (Palatini) variables when a
world metric and a symmetric world connection are considered on the
same footing. To compare the SEM flows in this model with those in
the Einsten's gravitational theory, we restrict our attention
the Hilbert-Einstein Lagrangian density in the Palatini variables. The
corresponding Euler-Lagrange equations
are well-known to be equivalent to the Einstein's equations \cite{bor}.

Let $LX\to X$ be the principal bundle of linear frames in the
tangent spaces to $X$. Its structure group is $GL^+(4,{\bf R})$.
The world connections are associated with the principal connections on the
principal bundle $LX\to X$.
 Hence, there is the 1:1 correspondence between the
world connections and the global sections of the quotient bundle
\begin{equation}
C_w=J^1LX/GL^+(4,{\bf R}). \label{251}
\end{equation}
With
respect to a holonomic atlas,  the bundle $C_w$ is coordinatized by
$(x^\la, k^\al{}_{\bt\la})$ so that, for any section $K$ of $C_w$,
$$
K^\al{}_{\bt\la}= k^\al{}_{\bt\la}\circ K
$$
are the coefficients of the linear connection
$$
K=dx^\la\otimes (\frac{\dr}{\dr x^\la} + K^\al{}_{\bt\la}\dot x_\al
\frac{\dr}{\dr \dot x_\bt})
$$
on $T^*X$.
In this
Section, we restrict ourselves to symmetric world connections.

The bundle $C_w$ (\ref{251}) admits the canonical splitting
$$
C_w=C_-\oplus C_+,
$$
$$
k^\al{}_{\bt\la}=k^\al{}_{[\bt\la]} +k^\al{}_{(\bt\la)},
$$
where
$$
C_-= \op\w^2T^*X\otimes TX
$$
is the bundle of torsion soldering forms and
$C_+\to X$ is the affine bundle modelled on the vector bundle
$$
\op\vee^2T^*X\otimes TX.
$$
Sections of the bundle $C_+\to X$ are symmetric world connections. This
bundle is coordinatized by $(x^\la, k^\al{}_{\bt\la})$ where $
k^\al{}_{\bt\la}=  k^\al{}_{\la\bt}$.

The total configuration space of the Palatini gravitational model is
\begin{equation}
J^1(\Si_g\op\times_X C_+). \label{C133}
\end{equation}
It is coordinatized by
$$
(x^\la, g^{\al\bt}, k^\al{}_{\bt\la}, g^{\al\bt}{}_\m, k^\al{}_{\bt\la\m}).
$$

On the configuration space (\ref{C133}), the  Hilbert-Einstein  Lagrangian
density reads
\begin{equation}
L_{\rm HE}=-\frac{1}{2\kp}g^{\bt\la}R^\al{}_{\bt\al\la}
\sqrt{-g}\omega,\label{C134}
\end{equation}
\[
R^\al{}_{\bt\n\la}=k^\al{}_{\bt\la\n}-
k^\al{}_{\bt\n\la}+k^\al{}_{\ve\n}k^\ve{}_{\bt\la}-k^\al{}_{\ve\la}
k^\ve{}_{\bt\n}.
\]
It is of the order zero with respect to the metric fields
$g^{\al\bt}$  and of the first order with respect to the coordinates
$k^\al{}_{\bt\la}$.

We consider the Palatini gravitational model in the presence of the Proca
fields.
 The total Lagrangian density $L$ on the
jet manifold $J^1T$ where
$$
T=\Si_g\op\times_X C_+\op\times_XT^*X
$$
is the sum
\begin{equation}
L= L_{\rm HE} +L_{\rm P}\label{C140}
\end{equation}
of the Hilbert-Einstein Lagrangian density  (\ref{C134}) and the
Lagrangian density of Proca fields  (\ref{C128}).

The associated Poincar\'e-Cartan form on the jet manifold $J^1T$
is the sum
$$
\Xi_L =\Xi_{\rm HE} +\Xi_{\rm P}
$$
of the form
$$
\Xi_{\rm HE} =-\frac1{2\kp}\sqrt{-g}g^{\bt\la}R^\al{}_{\bt\al\la}\om +
 \pi_\al{}^{\bt\n\la}\wh dk^\al{}_{\bt\n}\w\om_\la,
$$
\begin{equation}
\pi_\al{}^{\bt\n\la} =\frac{1}{2\kp}(\dl^\n_\al
g^{\bt\la}-\dl^\la_\al g^{\bt\n})\sqrt{-g},\label{C135}
\end{equation}
and the form $\Xi_{\rm P}$ (\ref{C137}). The
Euler-Lagrange operator corresponding to the Lagrangian density (\ref{C140})
is the sum
$$
\cE_L=\cE_{\rm P} +\cE_K +\cE_g
$$
 of the Euler-Lagrange operator for the Proca field $\cE_{\rm P}$
(\ref{C129}), for the
symmetric connection
\begin{equation}
\cE_K = \frac1{2\kp}[(\sqrt{-g}g^{\bt\nu})_{;\al}
-(\sqrt{-g}\dl^\nu_\al g^{\bt\la})_{;\la}] dk^\al{}_{\bt\nu}\w\om =
\dl_\al{}^{\bt\nu}\cL dk^\al{}_{\bt\nu}\w\om  \label{C143}
\end{equation}
where
$$
g^{\al\bt}{}_{;\la}= \wh\dr_\la g^{\al\bt} + k^\al{}_{\m\la}g^{\m\bt} +
k^\bt{}_{\m\la}g^{\al\m},
$$
and
for the metric field
\begin{equation}
\cE_g=\frac12\sqrt{-g}(T_{\al\bt}+ t_{\al\bt}) dg^{\al\bt}\w\om
 =\dl_{\al\bt}\cL dg^{\al\bt}\w\om\label{C138}
\end{equation}
where
\begin{equation}
T_{\al\bt}= -\frac1\kp(R_{\al\bt}
-\frac12g_{\al\bt}R)
\label{C139}
\end{equation}
and $t_{\al\bt}$ is the metric energy-momentum tensor (94) of Proca
fields.
 One can think of $T_{\al\bt}$
(\ref{C139}) as being the metric energy-momentum tensor of symmetric world
connections.

The total Lagrangian density $L$ (\ref{C140}) of the Palatini gravitational
model is invariant under general covariant transformations of the bundle of
geometric objects
$$
T=\Si_g\op\times_X C_+\op\times_XT^*X.
$$
Therefore, the Lie derivative of $L$
by the jet lift
$j^1_0\wt\tau$ of
the vector field on $T$
\ben
&&\wt\tau =\tau^\la\dr_\la + (g^{\nu\bt}\dr_\nu\tau^\al
+g^{\al\nu}\dr_\nu\tau^\bt)\frac{\dr}{\dr g^{\al\bt}}\nonumber\\
&& \qquad +[\dr_\nu\tau^\al k^\nu{}_{\bt\m} - \dr_\bt\tau^\nu
k^\al{}_{\nu\m} - \dr_\m\tau^\nu
k^\al{}_{\bt\nu} -\dr_{\bt\m}\tau^\al]\frac{\dr}{\dr k^\al{}_{\bt\m}}
\nonumber\\
&& \qquad -\dr_\al\tau^\nu k_\nu \frac{\dr}{\dr k_\al} \label{C141}
\een
naturally induced by a vector field $\tau$ on $X$ vanishes.
Hence, we get the week conservation law (\ref{K4}) in the form
\be
&& 0\ap \wh\dr_\la [\pi_\al{}^{\bt\mu\la}(-\tau^\nu k^\al{}_{\bt\m\nu} +
\dr_\nu\tau^\al k^\nu{}_{\bt\m} - \dr_\bt\tau^\nu
k^\al{}_{\nu\m} - \dr_\m\tau^\nu
k^\al{}_{\bt\nu} -\dr_{\bt\m}\tau^\al)\\
&&\qquad + \pi^{\nu\la}(-\dr_\nu\tau^\m k_\m
-\tau^\m k_{\nu\m}) +\tau^\la\cL].
\ee
 This can be brought  into the
form (\ref{K1}) where as like as in the previous case the total superpotential
is the sum of the gravitational Komar superpotential (\ref{K2}) and the
superpotential (\ref{2C13}) of Proca fields.

\bigskip\bigskip

\noindent
{\Large \bf 16\,\,\, Energy-momentum superpotential of affine-metric gravity}
\bigskip\bigskip

Let us consider the affine-metric gravitational model where dynamic variables
are pseudo-Riemannian metrics and general
linear connections on $X$.

Note that, since the world
connections are the principal connections,
one may apply the standard procedure of
gauge theory in order to obtain the SEM conservation law (see Section 10).
However in this case,  the  nonholonomic gauge isomorphisms of the linear frame
bundle $LX$ and the associated bundles have to be considered \cite{heh}. The
canonical lift $\wt\tau$ of a vector field $\tau$ on the base $X$ onto the
bundles $\Si_g$ and $C_w$ does not correspond to these isomorphisms. One must
use a horizontal lift of $\tau$ by means of some connection on these bundles.
At the same time, we observe that the configuration space of gauge
gravitation potentials in the gauge gravitation theory itself reduces to the
configuration space of general linear connections \cite{cam3}.

The total configuration space of the
affine-metric gravity is
\begin{equation}
J^1Y=J^1(\Si_g\op\times_XC_w) \label{N33}
\end{equation}
coordinatized by
$$
(x^\la, g^{\al\bt}, k^\al{}_{\bt\la}, g^{\al\bt}{}_\m,
k^\al{}_{\bt\la\m}).
$$

We assume that a Lagrangian density $L_{am}$ of the affine-metric gravitation
theory on the configuration space (\ref{N33}) depends on a metric $g^{\al\bt}$
and the curvature
\[
R^\al{}_{\bt\n\la}=k^\al{}_{\bt\la\n}-
k^\al{}_{\bt\n\la}+k^\al{}_{\ve\n}k^\ve{}_{\bt\la}-k^\al{}_{\ve\la}
k^\ve{}_{\bt\n}.
\]
In this case, we have the relations
\ben
&&\frac{\dr\cL_{\rm am}}{k^\al{}_{\bt\nu}}=
\pi_\si{}^{\bt\nu\la}k^\si{}_{\al\la} -
\pi_\al{}^{\si\nu\la}k^\bt{}_{\si\la},\nonumber\\
&& \pi_\al{}^{\bt\nu\la}=\dr_\al{}^{\bt\nu\la}\cL_{\rm am} =
-\pi_\al{}^{\bt\la\nu}. \label{K300}
\een

Let the Lagrangian density $L_{am}$ be invariant under general covariant
transformations.
Given a vector field $\tau$ on $X$, its canonical lift onto the bundle
$\Si_g\times C_w$ reads
\ben
&&\wt\tau =\tau^\la\dr_\la + (g^{\nu\bt}\dr_\nu\tau^\al
+g^{\al\nu}\dr_\nu\tau^\bt)\frac{\dr}{\dr g^{\al\bt}}\nonumber\\
&& \qquad +[\dr_\nu\tau^\al k^\nu{}_{\bt\m} - \dr_\bt\tau^\nu
k^\al{}_{\nu\m} - \dr_\m\tau^\nu
k^\al{}_{\bt\nu} -\dr_{\bt\m}\tau^\al]\frac{\dr}{\dr k^\al{}_{\bt\m}}.
\label{C151}
\een
For the sake of simplicity, let us employ the compact notation
$$
\wt\tau =\tau^\la\dr_\la + (g^{\nu\bt}\dr_\nu\tau^\al
+g^{\al\nu}\dr_\nu\tau^\bt)\dr_{\al\bt} + (u^A{}_\al^\bt\dr_\bt\tau^\al
-u^A{}_\al^{\ve\bt}\dr_{\ve\bt}\tau^\al)\dr_A.
$$

Since the Lie derivative of $L_{am}$
by the jet lift
$j^1_0\wt\tau$ of the field $\wt\tau$ (\ref{C151})
is equal to zero, i.e.
\begin{equation}
\bL_{j^1_0\wt\tau}L_{am}=0, \label{K6}
\end{equation}
we have the weak conservation law
\begin{equation}
0\ap \wh\dr_\la[ \dr^\la_A\cL_{\rm am}(u^A{}_\al^\bt\dr_\bt\tau^\al
-u^A{}_\al^{\ve\bt}\dr_{\ve\bt}\tau^\al -y^A_\al\tau^\al) +\tau^\la\cL_{\rm
am}]
\label{K8}
\end{equation}
where
\be
&& \dr^\la_A\cL_{\rm am} u^A{}_\al^{\ve\bt}
=\pi_\al{}^{\ve\bt\la},\\
&& \dr^\ve_A\cL_{\rm am} u^A{}_\al^\bt = \pi_\al{}^{\g\mu\ve}k^\bt{}_{\g\m} -
 \pi_\si{}^{\bt\mu\ve}k^\si{}_{\al\m} - \pi_\si{}^{\g\bt\ve}k^\si{}_{\g\al} =
\dr_\al{}^{\bt\ve}\cL_{\rm am} - \pi_\si{}^{\g\bt\ve}k^\si{}_{\g\al}.
\ee

Due to the arbitrariness of the functions $\tau^\al$, the equality (\ref{K6})
implies the strong equality
\begin{equation}
\dl^\bt_\al\cL_{\rm am} + \sqrt{-g}T^\al_\bt + u^A{}_\al^\bt\dr_A\cL_{\rm am} +
\wh\dr_\m(u^A{}_\al^\bt)\dr_A^\m\cL_{\rm am} -
 y^A_\al\dr_A^\bt\cL_{\rm am} =0.\label{K9}
\end{equation}
One can think of
$$
\sqrt{-g}T^\al_\bt =2g^{\al\nu}\dr_{\nu\bt}\cL_{\rm am}
$$
as being the metric energy-momentum tensor of general linear connections.

Substituting the term
$y^A_\al\dr_A^\bt\cL_{\rm am}$ from the expression (\ref{K9})
into the conservation law (\ref{K8}), we bring the latter into the form
\begin{equation}
0\ap \wh\dr_\la[ -\sqrt{-g}T^\la_\al\tau^\al
+\dr^\la_A\cL_{\rm am}(u^A{}_\al^\bt\dr_\bt\tau^\al
-u^A{}_\al^{\ve\bt}\dr_{\ve\bt}\tau^\al) -
\dr_A\cL_{\rm am} u^A{}_\al^\la\tau^\al -
\dr^\m_A\cL_{\rm am}\wh\dr_\m(u^A{}_\al^\la)\tau^\al]. \label{K10}
\end{equation}
Let us separate the components of the Euler-Lagrange operator
$$
\cE_L = (\dl_{\al\bt}\cL_{\rm am} dg^{\al\bt}
+\dl_\al{}^{\g\m}\cL_{\rm am} dk^\al{}_{\g\m})\w\om  $$
in the expression (\ref{K10}). We get
\be
&& 0\ap  \wh\dr_\la[ \dr^\la_A\cL_{\rm am} u^A{}_\al^\m\dr_\m\tau^\al
 - \wh\dr_\m(\dr^\m_A\cL_{\rm am} u^A{}_\al^\la)\tau^\al
+\wh\dr_\m(\pi_\al{}^{\ve\m\la})\dr_\ve\tau^\al] +\\
&& \qquad
\wh\dr_\la[-2g^{\la\m}\tau^\al\dl_{\al\m}\cL_{\rm am} -u^A{}_\al^\la\tau^\al
\dl_A\cL_{\rm am}] -
\wh\dr_\la[\wh\dr_\m(\pi_\al{}^{\nu\m\la}\dr_\nu\tau^\al)]
\ee
and then
\be
&& 0\ap \wh\dr_\la[
 - \wh\dr_\m(\dr_\al{}^{\la\m}\cL_{\rm am})\tau^\al] +\\
&& \qquad
\wh\dr_\la[-2g^{\la\m}\tau^\al\dl_{\al\m}\cL_{\rm am}
-(k^\la{}_{\g\m}\dl_\al{}^{\g\m}\cL_{\rm am} -
k^\si{}_{\al\m}\dl_\si{}^{\la\m}\cL_{\rm am} -
k^\si{}_{\g\al}\dl_\si{}^{\g\la}\cL_{\rm am})\tau^\al +
\dl_\al{}^{\ve\la}\cL_{\rm am}\dr_\ve\tau^\al]\\
&& \qquad
- \wh\dr_\la[\wh\dr_\m(\pi_\al{}^{\nu\m\la}(D_\nu\tau^\al+
\Om^\al{}_{\nu\si}\tau^\si)].
 \ee
The final form of the conservation law (\ref{K8}) is
\ben
&& 0\ap \wh\dr_\la[-2g^{\la\m}\tau^\al\dl_{\al\m}\cL_{\rm am}
-(k^\la{}_{\g\m}\dl_\al{}^{\g\m}\cL_{\rm am} -
 k^\si{}_{\al\m}\dl_\si{}^{\la\m}\cL_{\rm am} -
k^\si{}_{\g\al}\dl_\si{}^{\g\la}\cL_{\rm am})\tau^\al +
\dl_\al{}^{\ve\la}\cL_{\rm am}\dr_\ve\tau^\al
\nonumber\\
&& \qquad
-\wh\dr_\m(\dl_\al{}^{\la\m}\cL_{\rm am})\tau^\al]
- \wh\dr_\la[\wh\dr_\m(\pi_\al{}^{\nu\m\la}(D_\nu\tau^\al+
\Om^\al{}_{\nu\si}\tau^\si)]. \label{K11}
 \een

It follows that the SEM conservation law in the affine-metric gravity is
reduced to the form (\ref{K1}) where $U$ is the generalized
Komar superpotential (\ref{K3}).

Let us now examine the total system of the affine-metric gravity
and the tensor fields described in Section 8, e.g., the Proca
fields. In the presence of a general linear connection, the Lagrangian
density $L_{\rm P}$ (\ref{C128}) is naturally generalized through covariant
derivatives and depends on the torsion:
$$
\cF_{\mu\nu} = k_{\nu\mu} -k_{\mu\nu} - Q^\si{}_{\nu\mu}k_\si.
$$
It is readily observed that, in this case, the superpotential term
(\ref{2C13}) in the energy-momentum flow of the Proca fields (95) is
eliminated due to the additional contribution
$$
 -\wh\dr_\m (\dr_\al{}^{\la\m}\cL_{\rm P}\tau^\al).
$$
Thus, the energy-momentum flow of the Proca fields comes to zero, and the
total energy-momentum flow of affine-metric gravity and Proca fields reduces
to the generalized Komar superpotential.

One can consider general linear connections and Proca fields in the
presence of a background world metric $g$
when the general covariant transformations are not exact. In this case, we
have the weak transformation law (\ref{2C11}) where the variational
derivatives $\dl_{\al\bt}\cL$  by the metric field fail to vanish.
Then, the total SEM flow takes the form
$$
T^\la = \sqrt{-g}(T^\la_\al +t^\la_\al)\tau^\al +
\wh\dr_\m(\pi_\al{}^{\nu\m\la}(D_\nu\tau^\al+ \Om^\al{}_{\nu\si}\tau^\si))
$$
and the SEM transformation law comes to the form of the covariant
conservation law
$$
 (T^\la_\m +t^\la_\m)_{;\la}\ap 0
$$
of the metric energy-momentum tensors of general linear connections and
of the Proca fields. Thus, we observe that the "hidden" non-superpotential
part of the energy-momentum flow appears if the invariance under general
covariant transformations is broken.

\bigskip\bigskip

\noindent
{\Large \bf 17\,\,\,  Lagrangian systems on composite bundles}
\bigskip\bigskip

The gauge gravitation theory exemplifies the model with spontaneous
breaking of space-time symmetries where the matter fermion fields admit only
Lorentz transformations. The geometric formulation of the gauge gravitation
theory calls into play the composite bundle picture.
As a consequence, we
get the modified covariant differential of fermion fields which depends on
derivatives of gravitational fields \cite{cam3}.

In the gauge gravitation theory, gravity is represented by pairs $(h,A_h)$
of gravitational fields $h$ and associated Lorentz connections
$A_h$ \cite{heh,sard92}. The Lorentz connection $A_h$ is usually identified
with both a connection on a world manifold $X$ and a spinor connection on the
the spinor bundle $S_h\to X$ whose sections describe Dirac fermion fields
$\psi_h$ in the presence of the gravitational field $h$. The problem arises
when Dirac fermion fields are described in the framework of the affine-metric
gravitation theory. In this case, the fact that a world connection is
some Lorentz connection may result from the field
equations, but it can not be assumed in advance. There are models where the
world connection is not a Lorentz connection \cite{heh}. Moreover, it may
happen that a world connection is the Lorentz connection with respect
to different gravitational fields \cite{art}. At the same time, a
Dirac fermion field can be regarded only in a pair $(h,\psi_h)$ with a certain
gravitational field $h$.

Indeed, one must define the representation
of cotangent vectors to $X$ by the Dirac's $\g$-matrices in order to
construct the Dirac operator.
Given a tetrad gravitational field $h(x)$, we have the representation
$$
\g_h: dx^\mu \mapsto  \wh dx^\mu =h^\mu_a\g^a.
$$
However, different gravitational fields $h$ and $h'$ yield the
nonequivalent representations $\g_h$ and $\g_{h'}$.

It follows that, fermion-gravitation pairs
$(h,\psi_h)$ are described by sections of the composite spinor bundle
\begin{equation}
S\to\Si\to X \label{L1}
\end{equation}
where  $\Si\to X$  is the bundle of gravitational fields $h$;
the components $h^a_\m$ of $h$ play the role of
parameter coordinates of $\Si$, besides the familiar world coordinates.
\cite{sard92,sard1}. In particular, every spinor bundle $S_h\to X$ is
isomorphic to the restriction of $S\to\Si$ to $h(X)\subset \Si$. Performing
this restriction, we arrive at the familiar case of a field model in the
presence of a gravitational field $h(x)$.

By a composite bundle is meant the composition
\begin{equation}
Y\to \Si\to X. \label{1.34}
\end{equation}
of a bundle $Y\to X$ denoted by $Y_\Si$ and a bundle $\Si\to X$.
It is coordinatized by
$( x^\la ,\si^m,y^i)$
where $(x^\m,\si^m)$ are coordinates  of
$\Si$ and $y^i$ are the fiber coordinates of $Y_\Si$.
We further assume that $\Si$ has global sections.

The application of composite bundles to field theory is
founded on the following \cite{sard9}. Given
a global section $h$ of $\Sigma$, the restriction $ Y_h$
of $Y_\Si$ to $h(X)$ is a subbundle
of $Y\to X$. There is the 1:1 correspondence between
the global sections $s_h$ of $Y_h$ and the global sections of
the composite bundle (\ref{1.34}) which cover $h$.
Therefore, one can think of sections $s_h$ of $Y_h$ as
describing fermion fields in the presence of a background parameter
field $h$, whereas sections
of the composite bundle $Y$ describe all the pairs $(s_h,h)$.
The configuration space of these pairs is the
first order jet manifold $J^1Y$ of the composite bundle $Y$.

The feature of the dynamics of field systems on composite bundles consists in
the following.
Every connection
$$
A_\Si=dx^\la\ot(\dr_\la+\wt A^i_\la\dr_i)
+d\si^m\ot(\dr_m+A^i_m\dr_i)
$$
 on the bundle $Y_\Si$  yields
the horizontal splitting
$$
VY=VY_\Si\op\oplus_Y (Y\op\times_\Si V\Si),
$$
\[\dot y^i\dr_i + \dot\si^m\dr_m=
(\dot y^i -A^i_m\dot\si^m)\dr_i + \dot\si^m(\dr_m+A^i_m\dr_i).\]
Using this splitting, one can construct
the first order differential operator
\ben
&&\wt D:J^1Y\to T^*X\op\ot_Y VY_\Si,\nonumber\\
&&\wt D=dx^\la\ot(y^i_\la-\wt A^i_\la -A^i_m\si^m_\la)\dr_i,\label{7.10}
\een
on the composite
bundle $Y$. This operator posesses the following property.

Given a global section $h$ of $\Si$, let $\G$ be a connection on $\Si$
whose integral section is $h$, that is, $\G\circ h = J^1h$.
It is readily observed that the differential
(\ref{7.10}) restricted to $J^1Y_h\subset J^1Y$ comes
to the familiar covariant
differential relative to the connection
$$
A_h=dx^\la\ot[\dr_\la+(A^i_m\dr_\la h^m +\wt A^i_\la)\dr_i]
$$
on $Y_h$.
Thus,
we may utilize $\wt D$ in order to construct a Lagrangian density
$$
L:J^1Y\op\to^{\wt D}T^*X\op\ot_YVY_\Si\to\op\w^nT^*X
$$
for sections of the composite bundle $Y$.

\bigskip\bigskip

\noindent
{\Large \bf 18\,\,\,  Composite spinor bundles in gravitation theory}
\bigskip\bigskip

Let us consider the gauge theory of gravity and fermion fields.

Let $LX$ be the principal bundle of oriented linear frames in tangent spaces to
$X$. In gravitation theory, its structure group
$GL^+(4,{\bf R})$
is reduced to the connected Lorentz group
$ L=SO(1,3).$
It means that there exists a reduced subbundle $L^hX$ of $LX$ whose
structure group is $L$.
In accordance with the well-known theorem \cite{kob}, there is
the 1:1 correspondence between the reduced $L$ subbundles $L^hX$ of
$LX$ and the global
sections $h$ of the quotient bundle (\ref{C126}).

Given a section $h$ of $\Si$, let $\Psi^h$ be an atlas of $LX$ such that the
corresponding local sections $z_\xi^h$ of $LX$ take their values into $L^hX$.
With respect to $\Psi^h$ and a
holonomic atlas $\Psi^T=\{\psi_\xi^T\}$ of $LX$, a gravitational field $h$
can be represented by a family of $GL_4$-valued tetrad functions
\begin{equation}
h_\xi=\psi^T_\xi\circ z^h_\xi,\qquad
dx^\la= h^\la_a(x)h^a. \label{L6}
\end{equation}

By the Lorentz connections $A_h$ associated with a gravitational field $h$
are meant the principal connections on the reduced subbundle $L^hX$ of $LX$.
They give rise to principal connections on $LX$ and to spinor connections on
the $L_s$-lift $P_h$ of $L^hX$.

There are different ways to introduce Dirac fermion fields. Here, we follow the
algebraic approach.
Given a Minkowski space $M$, let
$ Cl_{1,3}$ be the complex Clifford algebra generated by elements
of $M$. A spinor space $V$ is defined to be a
minimal left ideal of $ Cl_{1,3}$  on
which this algebra acts on the left. We have the representation
$$
\g: M\ot V \to V
$$
of elements of the Minkowski space $M\subset Cl_{1,3}$ by
Dirac's matrices $\g$ on $V$.

 Let us consider a bundle of complex Clifford algebras $ Cl_{1,3}$ over $X$
whose structure group is the Clifford group of invertible elements of $
Cl_{1,3}$. Its subbundles are both a spinor bundle $S_M\to X$ and the bundle
$Y_M\to X$ of Minkowski spaces of generating elements of  $ Cl_{1,3}$.
To describe Dirac fermion fields on a world manifold $X$, one must
require $Y_M$ to be isomorphic to the cotangent bundle $T^*X$
of $X$. It takes place if there exists a reduced $L$ subbundle $L^hX$ of
$LX$ such that
\[Y_M=(L^hX\times M)/L.\]
Then, the spinor bundle
\begin{equation}
S_M=S_h=(P_h\times V)/L_s\label{510}
\end{equation}
is associated with the $L_s$-lift $P_h$ of $L^hX$. In this case, there exists
the representation
\begin{equation}
\g_h: T^*X\ot S_h=(P_h\times (M\ot V))/L_s\to (P_h\times
\g(M\times V))/L_s=S_h \label{L4}
\end{equation}
of cotangent vectors to a world manifold $X$ by Dirac's $\g$-matrices
on elements of the spinor bundle $S_h$. As a shorthand, one can write
\[\wh dx^\la=\g_h(dx^\la)=h^\la_a(x)\g^a.\]

Given the representation (\ref{L4}), we shall say that sections of
the spinor bundle $S_h$ describe Dirac fermion fields in the presence of
the gravitational field $h$. Indeed,
let
$$
A_h =dx^\la\ot (\dr_\la +\frac12A^{ab}{}_\la
I_{ab}{}^A{}_B\psi^B\dr_A)
$$
be a principal connection on $S_h$. Given
the corresponding covariant differential $D$ and the
representation $\g_h$ (\ref{L4}), one can construct the Dirac
operator \begin{equation}
\cD_h=\g_h\circ D: J^1S_h\to T^*X\op\ot_{S_h}VS_h\to VS_h, \label{I13}
\end{equation}
\[\dot y^A\circ\cD_h=h^\la_a\g^{aA}{}_B(y^B_\la-\frac12A^{ab}{}_\la
I_{ab}{}^A{}_By^B)\]
on the spinor bundle $S_h$.

Different  gravitational fields $h$ and $h'$ define nonequivalent
representations $\gamma_h$ and $\gamma_{h'}$.
It follows that a Dirac fermion field must be regarded only in a pair with
a certain gravitational field. There is the 1:1 correspondence
between these pairs and sections of the composite spinor bundle (\ref{L1})
defined as follows.

Let us consider the L-principal bundle
\[LX_\Si:=LX\to\Si\]
where $\Si$ is the quotient bundle (\ref{C126}).
 Let $P_\Si$ be the $L_s$-principal lift  of $LX_\Si$ such that
\[P_\Si/L_s=\Si, \qquad LX_\Si=r(P_\Si).\]
In particular, there is the imbedding of the $L_s$-lift $P_h$ of $L^hX$
onto the restriction of $P_\Si$ to $h(X)$.

We define the composite spinor bundle (\ref{L1}) where
\[S_\Si= (P_\Si\times V)/L_s\] is
associated with the $L_s$-principal bundle $P_\Si$. It is readily observed
that, given a global section $h$ of $\Si$, the restriction of $S_\Si$ to
$h(X)$ is the spinor bundle $S_h$ (\ref{510}) whose sections describe Dirac
fermion fields in the presence of the gravitational field $h$.

Let us provide the principal bundle $LX$ with a holonomic atlas
$\{\psi^T_\xi, U_\xi\}$ and the principal bundles $P_\Si$ and $LX_\Si$
with associated atlases $\{z^s_\e, U_\e\}$ and $\{z_\e=r\circ z^s_\e\}$.
With respect to these atlases, the composite spinor bundle is endowed
with the bundle coordinates $(x^\la,\si_a^\m, \psi^A)$ where $(x^\la,
\si_a^\m)$ are coordinates of the bundle $\Si$ such that
$\si^\m_a$ are the matrix components of the group element
$(\psi^T_\xi\circ z_\e)(\si),$
$\si\in U_\e,\, \pi_{\Si X}(\si)\in U_\xi.$
Given a section $h$ of $\Si$, we have
$$
 (\si^\la_a\circ h)(x)= h^\la_a(x),
$$
where $h^\la_a(x)$ are the tetrad functions (\ref{L6}).

Let us consider the bundle of Minkowski spaces
\[(LX\times M)/L\to\Si\]
associated with the $L$-principal bundle $LX_\Si$. Since $LX_\Si$ is
trivial, it is isomorphic to the pullback $\Si\op\times_X T^*X$
which we denote by the same symbol $T^*X$. Then, one can define the
bundle morphism
\begin{equation}
\g_\Si: T^*X\op\ot_\Si S_\Si= (P_\Si\times (M\ot V))/L_s
\to (P_\Si\times\g(M\ot V))/L_s=S_\Si, \label{L7}
\end{equation}
\[\wh dx^\la=\g_\Si (dx^\la) =\si^\la_a\g^a,\]
over $\Si$. When restricted to $h(X)\subset \Si$,
the morphism (\ref{L7}) comes to the morphism $\g_h$
(\ref{L4}).
We use this morphism in order to construct the total Dirac
operator on the composite spinor bundle $S$ (\ref{L1}).

Let
$$
\wt A=dx^\la\ot (\dr_\la +\wt A^B_\la\dr_B) + d\si^\m_a\ot
(\dr^a_\m+A^B{}^a_\m\dr_B)
$$
be a principal connection on the bundle $S_\Si$ and $\wt D$ the corresponding
differential (\ref{7.10}). We have the
first order differential
operator
\[\cD=\g_\Si\circ\wt D:J^1S\to T^*X\op\ot_SVS_\Si\to VS_\Si,\]
\[\dot\psi^A\circ\cD=\si^\la_a\g^{aA}{}_B(\psi^B_\la-\wt A^B_\la -
A^B{}^a_\m\si^\m_{a\la}),\] on $S$.
One can think of it as being the total Dirac operator since, for every
section $h$, the restriction of $\cD$ to $J^1S_h\subset J^1S$ comes
to the Dirac operator $\cD_h$ (\ref{I13})
relative to the connection
\[A_h=dx^\la\ot[\dr_\la+(\wt A^B_\la+A^B{}^a_\m\dr_\la h^\m_a)\dr_B]
\]
on the bundle $S_h$.

In order to construct the differential $\wt D$ (\ref{7.10}) on $J^1S$
in explicit form, let us consider the principal connection on the bundle
$LX_\Si$ given by the local connection form
\ben
&& \wt A = (\wt A^{ab}{}_\m dx^\m+ A^{ab}{}^c_\m d\si^\m_c)\ot I_{ab},
\label{L10}\\
&&\wt A^{ab}{}_\m=\frac12 K^\n{}_{\la\m}\si^\la_c (\eta^{ca}\si^b_\n
-\eta^{cb}\si^a_\n ),\nonumber\\
&&A^{ab}{}^c_\m=\frac12(\eta^{ca}\si^b_\m -\eta^{cb}\si^a_\m),
\label{M4}
\een
where  $K$ is a general linear connection on $TX$ and (\ref{M4})
corresponds to the canonical left-invariant free-curvature connection on
the bundle
\[GL^+(4,\R)\to GL^+(4,\R)/L.\]
 Accordingly, the differential $\wt D$
relative to the connection (\ref{L10}) reads
\begin{equation}
\wt D =dx^\la\ot[\dr_\la -\frac12A^{ab}{}_\m^c (\si^\m_{c\la} +
K^\m{}_{\nu\la}  \si^\nu_c)I_{ab}{}^A{}_B\psi^B\dr_A]. \label{K104}
\end{equation}

Given a section $h$, the connection $\wt A$ (\ref{L10}) is reduced to the
Lorentz connection
\begin{equation}
\wt K^{ab}{}_\la = A^{ab}{}_\m^c (\dr_\la h^\m_c +
K^\m{}_{\nu\la}  h^\nu_c) \label{K102}
\end{equation}
on $L^hX$,  and the differential
(\ref{K104}) leads to the covariant derivative of fermion fields
(\ref{K101}).

Let us emphasize that
the connection (\ref{K102})
is not the connection
$$
K^k{}_{m\la}=h^k_\m(\dr_\la h^\m_m +K^\m{}_{\nu\la} h^\nu_m) = K^{ab}{}_\la
(\eta_{am}\dl^k_b -\eta_{bm}\dl^k_a)
$$
written with respect to the reference frame $h^a=h^a_\la dx^\la$, but there is
the relation
\begin{equation}
\wt K^{ab}{}_\la=\frac12(K^{ab}{}_\la -K^{ba}{}_\la).\label{K50}
\end{equation}
If $K$ is a Lorentz connection $A_h$, then the connection $\wt K$
(\ref{K102}) consists with $K$ itself.

We utilize the differential (\ref{K104}) in order to construct a Lagrangian
density of Dirac fermion fields. This Lagrangian density
is defined on the configuration space $J^1(S\op\oplus_\Si S^+)$ coordinatized
by
$$
( x^\m, \si^\m_a, \psi^A,\psi^+_A, \si^\m_{a\la}, \psi^A_\la, \psi^+_{A\la}).
$$
It reads
\ben
 &&L_\psi=\{\frac{i}2[ \psi^+_A(\g^0\g^\la)^A{}_B( \psi^B_\la -
\frac12A^{ab}{}_\m^c (\si^\m_{c\la} +
K^\m{}_{\nu\la}  \si^\nu_c)I_{ab}{}^B{}_C\psi^C) -\nonumber\\
&& \qquad ( \psi^+_{A\la}- \frac12A^{ab}{}_\m^c (\si^\m_{c\la} +
K^\m{}_{\nu\la}  \si^\nu_c)\psi^+_C
I^+_{ab}{}^C{}_A)(\g^0\g^\la)^A{}_B\psi^B]  -
m\psi^+_A(\g^0)^A{}_B\psi^B\}\si^{-1}\om, \label{230}
\een
\[\g^\m=\si^\m_a\g^a, \qquad\si=\det(\si^\m_a),\]
where
$$\psi^+_A(\g^0)^A{}_B\psi^B$$
 is the Lorentz invariant fiber metric in the
bundle $S\op\oplus_\Si S^+$ \cite{cra}.

One can easily verify that
\begin{equation}
\frac{\dr\cL_\psi}{\dr K^\m{}_{\nu\la}} +
\frac{\dr\cL_\psi}{\dr K^\m{}_{\la\nu}} =0. \label{2C14}
\end{equation}
Hence, the Lagrangian density (\ref{230}) depends on the torsion
of the general linear connection $K$ only. In particular, it follows that, if
$K$ is the Levi-Civita connection of a gravitational field $h(x)$,
after the substitution $\si^\nu_c=h^\nu_c(x)$,
the Lagrangian density (\ref{230})
comes to the familiar Lagrangian density of fermion fields in the Einstein's
gravitation theory.
\bigskip\bigskip

\noindent
{\Large \bf 19\,\,\,  Conservation laws in gauge gravitation theory}
\bigskip\bigskip

In accordance with the previous Sections, the total configuration
space of fermion fields and affine-metric gravity can be described by the
jet manifold $J^1Y$ of the product
\begin{equation}
Y=S\op\oplus_\Si S^+\op\times_\Si C_w \label{2C15}
\end{equation}
where
$C_w$
is the bundle of general linear connections  (\ref{251}) coordinatized by
$(x^\la, k^\m{}_{\nu\la})$.

The total Lagrangian density $L$ on this configuration space is the sum of the
Lagrangian density $L_{am}$ of affine-metric gravity
where variables $g^{\al\bt}$ are replaced
with $\si^\al_a\si^\bt_b\eta^{ab}$
and the
Lagrangian density of fermion fields $L_\psi$ (\ref{230}) where the
background general linear connection $K^\m{}_{\nu\la}$ is replaced with the
corresponding coordinates $k^\m{}_{\nu\la}$ of the bundle $C_w$.

The Lagrangian
density $L$ is constructed to be invariant under the 1-parameter groups of
gauge isomorphisms of the $L$-principal bundle $LX\to \Si$.
The corresponding vector fields on the bundle $Y$ (\ref{2C15})
read
\begin{equation}
u=\frac12\al^{ab}(x)[(\eta_{ac}\dl_b^d - \eta_{bc}\dl_a^d)\si^\m_d\dr^c_\m +
I_{ab}{}^A{}_B\psi^B\dr_A + I^+_{ab}{}^A{}_B\psi^+_A\dr^B] \label{K110}
\end{equation}
where $\al^{ab}(x)$ are local parameters of gauge transformations.
The Lie derivative of the Lagrangian density $L$ by the jet lift $j^1_0u$ of
the vector field (\ref{K110}) is equal to zero and the corresponding
weak conservation law
$$
0\ap \wh\dr_\la[\frac12\al^{ab}(\dr^{c\la}_\m\cL_\psi(\eta_{ac}\dl_b^d -
\eta_{bc}\dl_a^d)\si^\m_d + \dr^\la_A\cL_\psi I_{ab}{}^A{}_B\psi^B+
I^+_{ab}{}^A{}_B\psi^+_A\dr^{B\la}\cL_\psi)]
$$
(\ref{K4}) takes place.
However, it is easy to verify that
\begin{equation}
\dr^{c\la}_\m\cL_\psi(\eta_{ac}\dl_b^d -
\eta_{bc}\dl_a^d)\si^\m_d + \dr^\la_A\cL_\psi I_{ab}{}^A{}_B\psi^B +
I^+_{ab}{}^A{}_B\psi^+_A\dr^{B\la}\cL_\psi=0,
\label{K111}
\end{equation}
and so the conserved current is equal to zero.

Now, we investigate the SEM conservation law of Dirac fermion fields and
affine-metric gravity. Let $\tau$ be a vector field on $X$. Both the Lagrangian
density $L_\psi$ of fermion fields and the Lagrangian density $L_{am}$ of
affine
metric gravity are invariant under the (local) 1-parameter groups
of transformations associated with the local vector fields
\begin{equation}
\wt\tau = \tau^\m\dr_\m +
\dr_\nu\tau^\m \si^\nu_a\frac{\dr}{\dr \si^\m_a} +
(\dr_\nu\tau^\al k^\nu{}_{\bt\m} - \dr_\bt\tau^\nu
k^\al{}_{\nu\m} - \dr_\m\tau^\nu
k^\al{}_{\bt\nu} -\dr_{\bt\m}\tau^\al)\frac{\dr}{\dr k^\al{}_{\bt\m}}.
 \label{K105}
\end{equation}
Note that, under a gauge (Lorentz) transformation, the field (\ref{K105}) is
changed as
$$
\wt \tau'=\wt\tau + \frac12\tau^\m\dr_\m
(\al^{ab})[(\eta_{ac}\dl_b^d - \eta_{bc}\dl_a^d)\si^\m_d\dr^c_\m +
I_{ab}{}^A{}_B\psi^B\dr_A +I^+_{ab}{}^A{}_B\psi^+_A\dr^B],
$$
but in virtue of the relation (\ref{K111}), the additional term in $\wt \tau'$
does not contribute in the SEM conservation law.

For the sake of simplicity, let us employ the same compact notation
as in Section 16:
$$
\wt\tau =\tau^\m\dr_\m + \dr_\nu\tau^\m \si^\nu_a\frac{\dr}{\dr \si^\m_a} +
(u^A{}_\al^\bt\dr_\bt\tau^\al -u^A{}_\al^{\ve\bt}\dr_{\ve\bt}\tau^\al)\dr_A.
$$

Since the Lie derivative of $L$
by the jet lift
$j^1_0\wt\tau$ of the field $\wt\tau$ (\ref{K105})
is equal to zero, i.e.
\begin{equation}
\bL_{j^1_0\wt\tau}L=0, \label{K200}
\end{equation}
the weak conservation law
\ben
&&0\ap \wh\dr_\la[ \dr^\la_A\cL_{\rm am}(u^A{}_\al^\bt\dr_\bt\tau^\al
-u^A{}_\al^{\ve\bt}\dr_{\ve\bt}\tau^\al -y^A_\al\tau^\al) \nonumber\\
&& \qquad +\frac{\dr\cL_\psi}{\dr\si^\al_{c\la}} (\dr_\bt\tau^\al\si^\bt_c -
\si^\al_{c\m}\tau^\m) - \frac{\dr\cL_\psi}{\dr\psi^A_\la}\psi^A_\al\tau^\al -
\frac{\dr\cL_\psi}{\dr\psi^+_{A\la}}\psi^+_{A\al}\tau^\al
+\tau^\la\cL] \label{K400}
\een
takes place. We have the relations (\ref{K300}) and the relation
$$
\frac{\dr\cL_\psi}{\dr k^\m_{\nu\la}} =\frac{\dr\cL_\psi}{\dr\si^\m_{c\la}}
\si^\nu_c.
$$

Due to the arbitrariness of the functions $\tau^\al$, the equality
(\ref{K200}) implies the strong equality (\ref{K9}) where $\sqrt{-g}$ is
replaced by $2\si$ and in addition the strong equality
\begin{equation}
\dl_\al^\bt\cL_\psi +2\si t_\al^\bt + \frac{\dr\cL_\psi}{\dr\si^\al_{c\la}}
\si^\bt_{c\la} - \frac{\dr\cL_\psi}{\dr\si^\m_{c\bt}}\si^\m_{c\al} +
\dr_A\cL_\psi u^{A\bt}_\al -
\frac{\dr\cL_\psi}{\dr\psi^A_\bt}\psi^A_\al -
\frac{\dr\cL_\psi}{\dr\psi^+_{A\bt}}\psi^+_{A\al} \label{K301}
\end{equation}
where
$$
2\si t^\bt_\al =\si^\bt_a\frac{\dr\cL_\psi}{\dr\si^\al_a}.
$$

Substituting the term $y^A_\al\dr_A^\bt\cL_{\rm am}$
from the expression (\ref{K9})
and the term
$$
\frac{\dr\cL_\psi}{\dr\si^\m_{c\bt}}\si^\m_{c\al} +
\frac{\dr\cL_\psi}{\dr\psi^A_\bt}\psi^A_\al +
\frac{\dr\cL_\psi}{\dr\psi^+_{A\bt}}\psi^+_{A\al}
$$
from the expression (\ref{K301})
into the conservation law (\ref{K400}), we bring the latter into the form
\ben
&& 0\ap \wh\dr_\la[-\si^\la_a\tau^\al\dl^a_\al\cL
-(k^\la{}_{\g\m}\dl_\al{}^{\g\m}\cL_{\rm am} -
 k^\si{}_{\al\m}\dl_\si{}^{\la\m}\cL_{\rm am} -
k^\si{}_{\g\al}\dl_\si{}^{\g\la}\cL_{\rm am})\tau^\al +
\dl_\al{}^{\ve\la}\cL_{\rm am}\dr_\ve\tau^\al
\nonumber\\
&& \qquad
-\wh\dr_\m(\dl_\al{}^{\la\m}\cL_{\rm am})\tau^\al]
- \wh\dr_\la[\wh\dr_\m(\pi_\al{}^{\nu\m\la}(D_\nu\tau^\al+
\Om^\al{}_{\nu\si}\tau^\si)] \nonumber\\
&&\qquad + \wh\dr_\la[(\frac{\dr\cL_\psi}{\dr\si^\al_{a\m}}\si^\la_a +
\frac{\dr\cL_\psi}{\dr\si^\al_{a\la}}\si^\m_a)\dr_\m\tau^\al]. \label{K303}
 \een
In accordance with the relation (\ref{2C14}), the last term in the expression
(\ref{K303}) is equal to zero, i.e. fermion fields do not contribute to the
superpotential. The SEM conservation law (\ref{K400}) comes to the form
(\ref{K1}) where $U$ is the generalized  Komar superpotential (\ref{K3}).

We can thus conclude that the generalized Komar superpotential (\ref{K3})
occurs rather universally in different gravitational models.

\bigskip

\centerline{\bf Acknowledgement}
\medskip

The authors would like to thank Prof. L.Mangiarotti and Prof. Y.Obukhov for
valuable discussions.

\end{document}